\documentclass[a4paper,12pt]{article}
\usepackage[super,mcite]{achemso} % ACS bib style
\usepackage[width=17.8cm,height=26.5cm]{geometry}
\usepackage[T1]{fontenc}
\usepackage{textcomp}
\usepackage{mathptmx}

\usepackage{graphicx}

\begin{document}

\title{DNA adsorption at liquid/solid interfaces}
\author{\bf Carine Douarche$^{\dag\ddag\S}$, Robert Cort\`es$^{\dag}$, Steven J. Roser$^{||}$, Jean-Louis Sikorav$^{\perp}$,\\
\bf and Alan Braslau$^{*\P}$}
\date{22 September 2008, submitted to the Journal of Physical Chemistry B}
\maketitle

\begin{center}
\it
Physique de la Mati\`ere Condens\'ee, \'Ecole Polytechnique, CNRS, 91128 Palaiseau, France;
Institut de Recherche Interdisciplinaire, Cit\'e Scientifique, Avenue Poincar\'e BP 60069, 59652 Villeneuve d'Ascq Cedex, France;
Department of Chemistry, University of Bath, Bath, Avon, UK, BA2 7AY;
Service de Biologie Int\'egrative et de G\'en\'etique Mol\'eculaire;
Institut de Physique Th\'eorique, CNRS URA 2306;
Service de Physique de l'\'Etat Condens\'e, CNRS URA 2464, CEA/Saclay, 91191 Gif-sur-Yvette Cedex, France
\end{center}

\begin{flushleft}
$^*$To whom correspondance should be addressed. E-mail: alan.braslau@cea.fr.\\
$^{\dag}$Physique de la Mati\`ere Condens\'ee, \'Ecole Polytechnique\\
$^{\ddag}$Service de Biologie Int\'egrative et de G\'en\'etique Mol\'eculaire, CEA/Saclay\\
$^{\S}$Institut de Recherche Interdisciplinaire, Villeneuve d'Ascq\\
$^{||}$Department of Chemistry, University of Bath\\
$^{\perp}$Institut de Physique Th\'eorique, CEA/Saclay\\
$^{\P}$Service de Physique de l'\'Etat Condens\'e, CEA/Saclay\\
\end{flushleft}

\begin{abstract}
DNA adsorption on solid or liquid surfaces is a topic of broad fundamental and applied interest. Here we study by x-ray reflectivity the adsorption of monodisperse double-stranded DNA molecules a positively-charged surface, obtained through chemical grafting of a homogeneous organic monomolecular layer of N-(2-aminoethyl) dodecanamide on an oxide-free monocrystalline Si(111) wafer. The adsorbed dsDNA is found to embed into the soft monolayer which is deformed in the process. The surface coverage is very high and this adsorbed layer is expected to display 2D nematic ordering.
\end{abstract}

\section*{Introduction}

The study of DNA-interface interactions is a subject of general nature, as illustrated in Fig.~1. DNA interactions with surfaces can be repulsive leading to confinement or depletion\cite{Asakura1954,Zimmerman1993}. A repulsive interaction exists between negatively charged phosphates of DNA and phospholipids of membranes and contributes to enclose the genetic material within the cell\cite{Westheimer1987}. Depletion forces also confine DNA within viral capsids\cite{Odijk2004}. Repulsive interactions are a common requirement in micro and nanofluidics\cite{Squires2005}.

Attractive interactions lead to adsorption. Chemisorption is at the core of solid-phase synthesis of polynucleotides\cite{Merrifield1985}. The case of physisorption is also of wide technological interest and biological relevance. In hybridization techniques such as Southern blots, Northern blots, or biochips, nucleic acids are immobilized on surfaces, either through chemisorption or physisorption. Chromatographic techniques rely on a reversible immobilization of DNA. Hybridization technology is, in fact, a particular type of affinity chromatography\cite{Bautz1962,Cuatrecasas1968,Schott1984}. The imaging of DNA macromolecules by electron microscopy\cite{Lang1964,Vollenweider1975}, scanning tunnelling microscopy (STM) and atomic force microscopy (AFM)\cite{Hansma1993} require their adsorption at the air-liquid or liquid-solid interface. DNA adsorption at the air-liquid interface is coupled to DNA aggregation\cite{Douarche2007}, establishing a link between the fields of DNA adsorption and DNA condensation\cite{Bloomfield1997}. The development of single-molecule studies of DNA relies on tethering to a solid surface (microscope slide or bead). DNA adsorption is also a essential aspect of DNA combing and mapping\cite{Allemand1997,Jing1998}.

The mobility of adsorbates, which is a central issue of surface chemistry, is also of great importance for nucleic acids. Under weak yet irreversible adsorption, the DNA is free to rearrange to its equilibrium configuration; under strong adsorption, the chains may, nevertheless, maintain some mobility, in particular when the interface itself is fluid\cite{Lang1964,Frontali1979,Joanicot1987,Maier2000,Goldar2004}, or be free to roll on the surface\cite{Rhodes1980}. This can lead to an increase in the efficiency of a diffusional search through a reduction of dimensionality\cite{Adam1968}. For example, the diffusion of proteins on the surface of DNA (or {\it vice versa}) can greatly increase the rate of sequence recognition\cite{Berg1985,Wang2006}.

The notions addressed here are that of depletion versus adsorption, of confinement, of chemisorption versus physisorption, and of immobilization versus surface diffusion. These notions are associated with the most fundamental concepts of surface science, heterogeneous catalysis and enzymology.

\section*{Experimental Methods and Results}

In this context, we report one particular investigation using x-ray reflectivity to characterize the process of DNA adsorption at a liquid/solid interface. We have chosen single-crystal silicon as a substrate material since its topography is well-defined at the atomic scale, its surface chemistry is well-controlled, and for its interest of use in biochip technologies.

\subsection*{Surface functionalization}

Starting from an etched Si(111) wafer, we have assembled through a multi-step chemical process a covalently-bonded positively charged molecular monolayer\cite{DouarcheThesis2007}, as illustrated in Fig.~2(a) and 2(b). The native silicon oxide layer was removed by etching in an ammonium fluoride solution\cite{Allongue2000}. An ester-terminated alkyl chain, ethyl undecylenate [CH$_2$=CH--(CH$_2$)$_8$--CO--O--C$_2$H$_5$], was then covalently and densely attached to the regular silicon crystal surface. This molecular layer was hydrolyzed in hydrochloric acid and activated using N-hydroxysuccinimide (NHS -- C$_4$H$_5$NO$_3$) before coupling to ethylene-diamine (NH$_2$--CH$_2$--CH$_2$--NH$_2$), thus obtaining a terminal positively-charged group. The grafted molecule has the structure of N-(2-aminoethyl) dodecanamide (CAS 10138-02-0), where the terminal methyl group is replaced by a surface silicon atom of the monocrystalline substrate.

The surfaces were then characterized by Fourier transformed infrared (FTIR) spectroscopy and by AFM\cite{DouarcheThesis2007}. The images obtained by AFM in air of the bare H-Si(111) surface before [Fig.~2(c)] and after functionalization with the dense organic layer [Fig.~2(d)] are comparable: the atomic flatness of the terraced structure of the single-crystal silicon surface is preserved in the chemical grafting process. FTIR spectroscopy studies\cite{DouarcheThesis2007} show that the functionalized surface is chemically homogeneous with a surface coverage of $(3.1\pm 0.3)\times 10^{14}\ \mathrm{molecules}/\mathrm{cm}^2$ which corresponds to 40$\pm$4{\%} of the available Si-H binding sites. The inverse of this surface density yields an area occupied on average per grafted molecule ($0.32\pm 0.03\ \mathrm{nm}^2/\mathrm{molecule}$) which is that of a tilted, condensed monolayer phase (and about 50{\%} greater than that of the perpendicular area of a densely-packed alkyl chain). The grafted surfaces remain oxide-free due to the very high surface-density of the organic monomolecular layer.
  
\subsection*{X-ray reflectivity data analysis}

The surfaces as prepared were further characterized by synchrotron x-ray reflectivity at the European Synchrotron Radiation Facility at Grenoble, France. Using a procedure of contrast variation common for neutron scattering, the samples were measured both in air and under aqueous buffer and the data analyzed conjointly (Fig.~3, see reference~\cite{DouarcheJCP} for details). The organic layer can be observed with great contrast with respect to air resulting in a strong Kiessig fringe\cite{Kiessig1931b} as seen in Fig.~3(a) (black circles); under buffer, the contrast is diminished (red triangles), but the reflectivity here is more sensitive to the positively-charged diamine head-groups than when recorded under air. The measured reflectivity can be described by a model density-profile in the direction normal to the surface, as shown in Fig.~3(b): the black dot-dashed line is the profile with air and the red dashed line is the profile under buffer. The thickness of the covalently bonded, dense, monomolecular layer is measured to be $1.504 \pm 0.114\ \mathrm{nm}$. This layer can be divided into an alkyl-chain region of density $\sim 1\ \mathrm{g}/\mathrm{cm}^3$ and a hydrophilic head-group region of higher density. The root-mean-squared roughness, averaged over the $\sim 1\ \mathrm{cm}^2$ illuminated surface, is measured to be $0.3542 \pm 0.0039\ \mathrm{nm}$, somewhat larger than the ideal case, indicating some disorder of the functional head-groups. As the data measured under air and under buffer are considered conjointly, the model density profile is constrained and one further parameter can be extracted, namely the integral density or, dividing by the layer thickness, the surface density; The x-ray reflectivity measurements yield $(1.54\pm 0.15)\times 10^{-7} \mathrm{g}/\mathrm{cm}^2 = (3.40\pm 0.33)\times 10^{14}\ \mathrm{molecules}/\mathrm{cm}^2$, in excellent agreement with the FTIR estimation.

\subsection*{DNA adsorption}

These homogeneous, positively-charged functionalized surfaces have been used to probe DNA adsorption. Monodisperse, double-stranded DNA (dsDNA) molecules were obtained by PCR amplification from a 294 base-pair sequence (nucleotides 175-470) of the 5386~bp genome from the phage $\phi$X174, used as a template. The DNA was studied under a 1~mM sodium phosphate buffer solution (NaH$_2$PO$_4$ + Na$_2$HPO$_4$, pH~7.2). At this low salt concentration, the Debye length is about 10~nm and the contour length of the monodisperse dsDNA molecules employed\cite{DouarcheJCP}, $\ell = 294\ \mathrm{bp}\cdot 0.34\ \mathrm{nm}/\mathrm{bp} = 100\ \mathrm{nm}$, is close to the persistence length of dsDNA\cite{Baumann1997}. Such molecules, when adsorbed, loose one degree of freedom with the result that the persistence length is doubled\cite{Frontali1979}. Therefore, we must consider here dsDNA molecules that are at or near the rod limit\cite{Odijk1977}, (semi-flexible) cylinders of diameter $\oslash = 2\ \mathrm{nm}$, length $L \approx \ell$ (having an aspect ratio $L/\oslash = 50$) and, for planar adsorption, covering an area of $\ell \cdot \oslash = 200\ \textrm{nm}^2$. The surface density of dsDNA necessary for complete, saturated monomolecular coverage of the surface can be estimated to be $(294\ \mathrm{bp}\cdot 660\ \mathrm{g}\cdot \mathrm{mole}^{-1}\mathrm{bp}^{-1}/N_A) / (200\ \mathrm{nm}^2) = 161\ \mathrm{ng}/\mathrm{cm}^2$. As the total surface area of the functionalized silicon wafer in the sample cell is $3\ \mathrm{cm} \times 2.4\ \mathrm{cm}$, this coverage would correspond to a total adsorbed mass of $m_{\mathrm{sat}} = 1.16\ \mu\mathrm{g}$. One can also define a second quantity of much lower surface coverage, $C^{\star} = \ell\cdot\oslash/\pi(\ell/2)^2 = 4\oslash / \pi\ell = 0.026$, or $4.1\ \mathrm{ng}/\mathrm{cm}^2$ in terms of surface mass density, below which the rods could be independently adsorbed and isotropically oriented in the surface plane. Above this concentration, an adsorbed monolayer of dsDNA will exhibit locally some degree of 2D nematic order (to be discussed below).

A solution containing $1.16\ \mu\mathrm{g}$ of dsDNA ($1\times m_{\mathrm{sat}}$) was added to the sample cell; the adsorption of the dsDNA chains remained undetected by x-ray reflectivity under repeated measurements over a many hours of incubation. The quantity of dsDNA in the solution was increased by an order of magnitude ($10\times m_{\mathrm{sat}}$) and, after about 6~hours of identical, reproducible results, we then measured a very different (and reproducible) reflectivity profile as shown by the blue squares in Fig.~3(a) and the blue solid line in Fig.~3(b). This new density profile remained unchanged 1) over time, 2) upon increasing bulk concentration of dsDNA (up to a $60\times m_{\mathrm{sat}}$), 3) after rinsing with the buffer solution, and, finally, 4) even under buffer of increasingly high monovalent ionic concentrations (up to 500~mM NaCl). Extensive tests were also performed\cite{DouarcheJCP} to assure that the observed change in the density profile was not due to radiation damage to the functionalized substrate. An analysis\cite{DouarcheJCP} of the resulting density profile (blue solid line in Fig.~3) shows an increase in the surface mass-density but no significant increase in the surface layer thickness. The integrated increase in surface mass density incorporated into the grafted layer is measured to be $(35\pm 16)\ \mathrm{ng}/\mathrm{cm}^2$. This quantity is close to $10\times$ that of the overlap quantity $C^{\star}$ ($4.1\ \mathrm{ng}/\mathrm{cm}^2$), yet less than the estimated density of complete coverage ($161\ \mathrm{ng}/\mathrm{cm}^2$).

\section*{Discussion}

Since the volumic mass density of dsDNA in water is very close to $1\ \mathrm{g}/\mathrm{cm}^2$, the x-ray reflectivity measurements reported here are insensitive to a {\em disordered} layer of dsDNA adsorbed {\em onto} the functionalized surface, in agreement with the absence of observed electron density beyond the grafted layer thickness. A Mikado (pick-up sticks) game-like isotropic dispersion would remain mostly undetected by x-ray reflectivity and cannot account for the measured increase in surface mass-density. This image is pertinent due to the rigidity of the rod-like dsDNA. Furthermore, such a disordered configuration is unlikely as DNA crossings are energetically unfavorable\cite{Brenner1974} at low ionic strengths (here 1~mM sodium phosphate).

The high coverage observed suggests a local, 2D nematic order\cite{Ubbink2004}, such as is drawn in Fig.~4. At the beginning of the adsorption process, the surface coverage is low, events are uncorrelated and DNA molecules stick to the surface at random, as stated above. As the surface coverage approaches the overlap quantity $C^{\star}$, impinging molecules will be directed to more parallel orientations by the electrostatic repulsion of the already adsorbed molecules, yielding ordered domains with a characteristic size given by the length of the rod-like dsDNA molecules (100~nm). Such a process can generate a 2D nematic order even if the adsorbed chains are immobilized. If, however, the adsorbed molecules are able to rearrange, either through sliding or rolling, then the extent of these ordered domains could grow through some sort of annealing process. Note, however, that the overlap condition suggests a domain size that fortuitously corresponds to the width of the atomic terraces ($\sim$90~nm); thus, the question of equilibrium of the strongly adsorbed molecules is not addressed here.

The measured increased surface mass-density corresponds to a partial penetration of the monomolecular layer of adsorbed dsDNA molecules {\em into} the soft grafted layer, bringing additional mass density to this region. The degree of penetration is a function of the surface coverage of dsDNA. At one extreme, if the adsorbed molecules were to penetrate into the layer (thickness 1.5~nm) all the way down to touching the hard silicon substrate [Fig.~4(a)] through a splay deformation of the grafted chain molecules, 80{\%} of their mass would be incorporated into the monolayer\footnote{The fraction of a cylinder of diameter 2~nm buried in a layer of thickness 1.5~nm ($1.5\times$ the radius) is $1 - [1/3 - \sqrt{3/4}/(2\pi)] = 0.8045$.}. This model yields an estimate of the real molecular surface coverage of $(35\pm 16)/0.80 = (43\pm 20)\ \mathrm{ng}/\mathrm{cm}^2$ or a fractional surface coverage of $(43 \pm 20)\ \mathrm{ng}\cdot\mathrm{cm}^{-2}/161\ \mathrm{ng}\cdot\mathrm{cm}^{-2} = 0.27\pm 0.12$. In this case, the estimated average distance between DNA double-helical chains would be $2\ \mathrm{nm}/0.27 = 7.4\ \mathrm{nm}$ (between 5.1~nm and 14~nm due to the uncertainty of the measurement). However, this situation of such deep incorporation is unlikely because of the covalent grafting of the alkyl-chain monolayer. The other extreme, that of total monomolecular dsDNA coverage [Fig.~4(c)], is also unlikely as it implies an intimate contact between ordered, parallel, dsDNA chains with unfavorable energetics\cite{Rau1984}. The effective penetration into the layer in this case would be less, as the measured value of incorporated mass density corresponds to $(35\pm 16)\ \mathrm{ng}\cdot\mathrm{cm}^{-2} / 161\ \mathrm{ng}\cdot\mathrm{cm}^{-2} = 0.22\pm 0.1$ of the theoretical saturated surface mass density: a simple calculation of the sagitta of the cylindrical cross-section of the adsorbed dsDNA molecule estimates a penetration depth of 0.54~nm. We conclude that the measurements yield an estimated incrustation depth that lies in the range of 0.54--1.5~nm for a surface coverage in the range of 100--27\%, corresponding to an average distance between molecular axes in the range of 2--7.4~nm. A similar result is observed by AFM for dsDNA condensation on cationic lipid membranes, with a measured chain-chain separation $\sim 5\ \mathrm{nm}$\cite{Fang1997} that is compatible with the range estimated here.

\section*{Conclusions}

The single-crystal substrates studied here were oxide free and chemically homogeneous with a dense, monomolecular organic layer. Such a layer, covalently attached to the monocrystalline substrate remains soft and deformable, even for the relatively short alkyl chain lengths used here. We note that this situation of polyelectrolyte adsorption at a (crystalline) hard wall covered by a dense, grafted, yet deformable monomolecular layer has not been addressed theoretically, contrary to that of a hard wall alone\cite{Ubbink2004} or on a deformable membrane\cite{Podgornik1997}. The conformation of DNA upon adsorption at solid interfaces as well as its mobility depends on the details of the interactions present and the structure of the solid interface. These details are generally not well known or well characterized.

The present system is most promising for the development of nucleic acid biochip hybridization technologies\cite{Squires2008}. These devices are commonly designed such that the immobilized DNA probes extend into the solution and away from the surface\cite{Southern1999}. Hybridization with the targets then takes place directly from the bulk solution. This mechanism is know as an Eley-Rideal\cite{Eley1940,Eley1941} reaction in surface chemistry. Alternatively, one can seek to increase the efficiency of this process by the adsorption of the complementary strands (both targets and ``immobilized'' probes) which could then meet through a two-dimensional diffusion\cite{Adam1968,Chan1995,Goldar2004}. Such a reaction between two adsorbates is called a Langmuir-Hinshelwood\cite{Langmuir1922,Hinshelwood1926} mechanism. The application of this scheme will require well-defined, functionalized surfaces. In contrast with approaches involving thiol grafting on gold layers or silanation on silicon-oxide layers, the use of oxide-free silicon surfaces is particularly pertinent as they can be efficiently charge coupled to grafted or adsorbed DNA molecules.

\section*{Acknowledgements}

The authors wish to acknowledge A.~Goldar (CEA/Saclay) for inspiration, constant support and detailed discussions. C.~D. benefitted from financial support by the CNRS/IRI, Villeneuve d'Ascq (R.~Boukherroub) for her thesis work. We thank C. Henry~de~Villeneuve (PMC, \'Ecole polytechnique) for participation in the synchrotron experiments, performed at the ID10B beamline of the European Synchrotron Radiation Facility, Grenoble, France with the skillful assistance of A.~Vorobiev and O.~Konovalov. We also thank the anonymous referee for requesting clarifications on the adsorption process.

%\bibliography{../DNAonSi}
\ifx\mcitethebibliography\mciteundefinedmacro
\PackageError{achemsoM.bst}{mciteplus.sty has not been loaded}
{This bibstyle requires the use of the mciteplus package.}\fi

\section*{Figures}

\begin{description}

\item[Figure 1.] Surface science of DNA.\\
\includegraphics{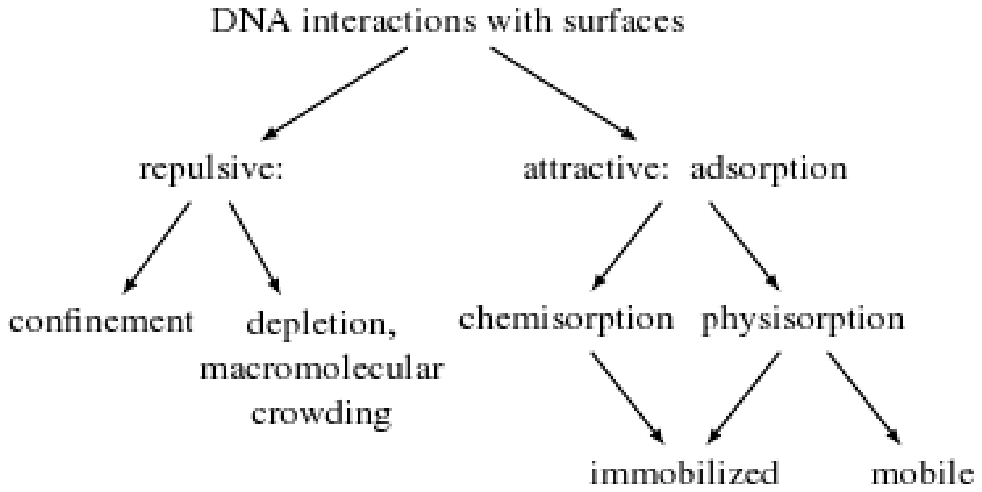}

\newpage
\item[Figure 2.] Si(111). (a) Molecular model of the atomically-stepped surface. The terrace width is $d_{(111)}/\tan\alpha = 90\ \mathrm{nm}$ where $d_{(111)} = 0.314\ \mathrm{nm}$ is the Si(111) layer spacing and $\alpha$ is the surface miscut angle (0.2$^\circ$, drawn here at 5$^\circ$ for illustration) of the surface normal $\hat{z}$ in the $[11\overline{2}]$ direction. The surface density of Si-H binding sites is $7.8\times 10^{14}\ \mathrm{cm}^{-2}$. (b) Molecular model of the grafted monomolecular surface layer. (c) AFM image of the bare H-Si(111) surface after etching in NH$_4$F; image scale: 2~$\mu$m. (d) AFM image of the Si(111)-H functionalized with the molecular monolayer; image scales (inset): 2~$\mu$m, 1~$\mu$m and 500~nm.\\
\includegraphics{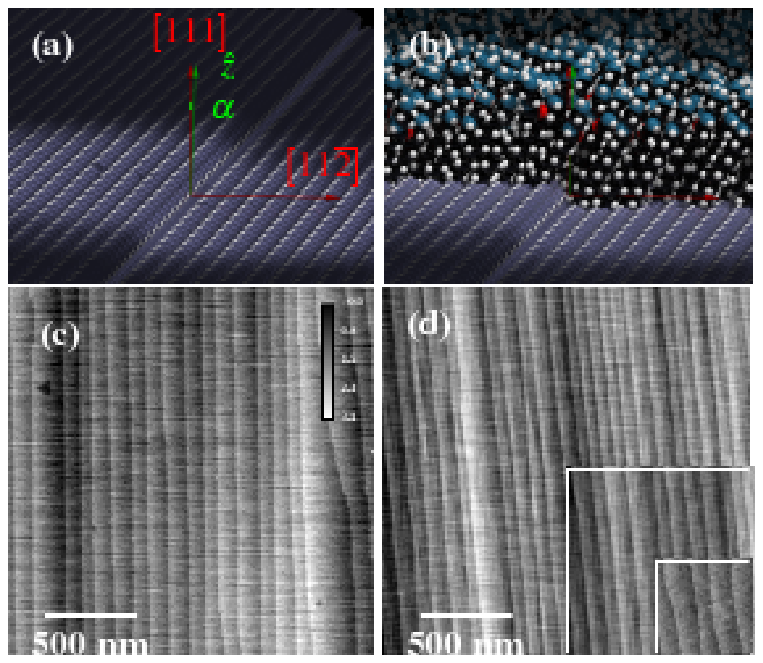}

\newpage
\item[Figure 3.] X-ray reflectivity and model density profiles. (a) Normalized specularly-reflected intensity as a function of wave-vector transfer $Q_z = (4\pi/\lambda)\sin\theta$; $\lambda$ is the x-ray wavelength and $\theta$ is the grazing angle of incidence. Black circles: functionalized substrate measured in air; Red triangles: under buffer; Blue squares: after adsorption of DNA. The incident beam footprint overflows the substrate for data at $q_z < 1\ \mathrm{nm}^{-1}$ (gray region). Insert: kinematics of the measurement; (b) Model density profiles $\rho(z)$. Black dot-dashed line: functionalized substrate measured in air; Red dashed line: under buffer; Blue solid line: after adsorption of DNA. A model of the grafted molecule is drawn for comparison.\\
\includegraphics{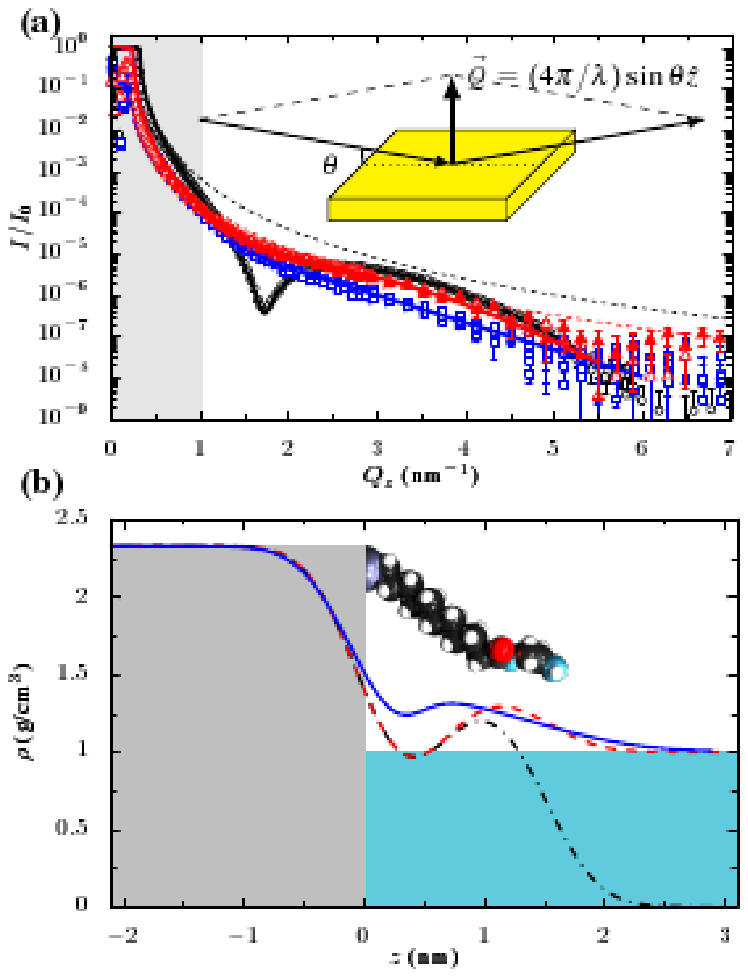}

\newpage
\item[Figure 4.] Embedding of the adsorbed DNA molecules. Two extreme cases: (a) maximal embedding. The real coverage in this case is 27{\%} and the average lateral separation is 7.4~nm; (c) maximal (100{\%}) coverage. The embedding depth is $(0.54\pm 0.17)\ \mathrm{nm}$ for a lateral separation of 2~nm. Intermediate case: (b) 50{\%} coverage. The embedding depth for this coverage is 1~nm and the lateral separation is 4~nm. The grafted molecular monolayer [see Fig.~2(b)] is here schematized as a solid, blue layer for clarity.\\
\includegraphics{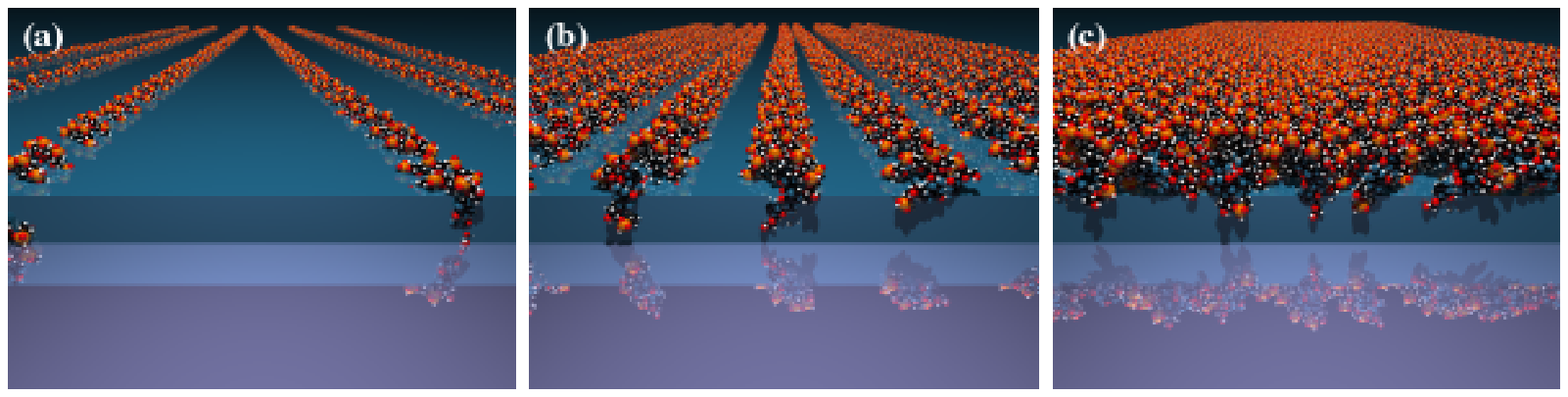}

\end{description}

\end{document}